\documentclass[english,aps,prb,10pt,twocolumn,superscriptaddress,showpacs]{revtex4}
\usepackage[T1]{fontenc}
\usepackage[latin9]{inputenc}
\usepackage{tabularx}
\usepackage{graphicx,wrapfig}
\usepackage{babel}
\usepackage{layouts}
\usepackage{graphicx}
\usepackage{epstopdf}
\usepackage{caption}

\usepackage{subfigure}
\usepackage{dcolumn}
\usepackage{bm}
\usepackage{amsmath}
\usepackage{comment}
\usepackage{bbold}
\DeclareMathOperator{\di}{d\!}

\usepackage{color}

\makeatletter
\@ifundefined{textcolor}{}
{%
 \definecolor{BLACK}{gray}{0}
 \definecolor{WHITE}{gray}{1}
 \definecolor{RED}{rgb}{1,0,0}
 \definecolor{GREEN}{rgb}{0,1,0}
 \definecolor{BLUE}{rgb}{0,0,1}
 \definecolor{CYAN}{cmyk}{1,0,0,0}
 \definecolor{MAGENTA}{cmyk}{0,1,0,0}
 \definecolor{YELLOW}{cmyk}{0,0,1,0}
 }

\newcommand{\R} {\mbox{Re}\,}
\newcommand{\I} {\mbox{Im}\,}

\newcommand{\bea}{\begin{eqnarray}}
\newcommand{\eea}{\end{eqnarray}}

\newcommand{\be}{\begin{equation}}
\newcommand{\ee}{\end{equation}}

\newcommand{\ket}[1]{\left|{#1}\right\rangle}
\newcommand{\bra}[1]{\left\langle{#1}\right|}

\makeatother

\begin{document}


\title{Cavity-coupled double-quantum dot at finite bias: analogy with lasers and beyond}


\author{Manas Kulkarni}
\affiliation{Department of Electrical Engineering, Princeton University,
Princeton, NJ 08544, USA}
\author{Ovidiu Cotlet}
\affiliation{Department of Physics, Princeton University,
Princeton, NJ 08544, USA}
\author{Hakan E. T\"{u}reci}
\affiliation{Department of Electrical Engineering, Princeton University,
Princeton, NJ 08544, USA}
\date{\today}

\begin{abstract}
We present a theoretical and experimental study of photonic and electronic transport properties of a voltage biased InAs semiconductor double quantum dot (DQD) that is dipole-coupled to a superconducting transmission line resonator. We obtain the Master equation for the reduced density matrix of the coupled system of cavity photons and DQD electrons accounting systematically for both the presence of phonons and the effect of leads at finite voltage bias. We subsequently derive analytical expressions for transmission, phase response, photon number and the non-equilibrium steady state electron current. We show that the coupled system under finite bias realizes an unconventional version of a single-atom laser and analyze the spectrum and the statistics of the photon flux leaving the cavity. In the transmission mode, the system behaves as a saturable single-atom amplifier for the incoming photon flux. Finally, we show that the back action of the photon emission on the steady-state current can be substantial. Our analytical results are compared to exact Master equation results establishing regimes of validity of various analytical models. We compare our findings to available experimental measurements.
%

\end{abstract}



\maketitle

\textit{Introduction - } Hybrid light-matter systems offer a unique platform to study non-equilibrium many-body phenomena. Here we study a hybrid system composed of a voltage-biased double quantum dot (DQD) coupled to a superconducting transmission line resonator \cite{pettaNature,frey,kontosnatcom,petta2014,kontos2013ar}. From the perspective of Cavity QED physics which studies fundamental light-matter interaction between a quantum emitter and the radiation field of an electromagnetic resonator, here we encounter a situation where the electronic excitations constituting the emitter cannot be treated as an isolated system. Many-body effects that derive from the hybridization of the DQD electrons with the lead electrons can lead to novel phenomena at the interface of quantum impurity physics and quantum optics. In particular, such a setting provides an accurate probe to study the non-equilibrium dynamics of standard quantum impurity systems. A case in point is recent theoretical \cite{het2011} and subsequent experimental work \cite{het2011nat} that studied the optical signatures of the quench dynamics of a quantum impurity in the Kondo regime. More recent work \cite{hetArxiv} addressed a strongly (optically) driven quantum impurity beyond the linear response regime and demonstrated the formation of a new quantum-correlated state characterized by the emergence of a secondary spin-screening cloud. 

Here we study a cavity coupled to a quantum impurity through which a steady-state current is driven. The quantum impurity is a DQD defined on a nanowire through electrostatic gating and held under a static source-drain voltage. Such a system acts as an amplifier for an incident microwave signal. 
Photon emission from this system has recently been experimentally analyzed \cite{petta2014} showing a gain as large as 15 in the cavity transmission. Earlier theoretical work has addressed statistics of photon emission \cite{xu2013quantum,xu2013full,schon2013,jin2011lasing} and intracavity photon number\cite{childress2004mesoscopic} but did not touch on the interplay of phonon and photon-induced interdot tunneling processes, which experiments of Ref. \onlinecite{petta2014} indicate to play an important role in determining the gain and the bandwidth of this cavity-DQD quantum amplifier. A systematic analysis assessing the role of photonic and phononic quantum environment in the non-equilibrium dynamics of this hybrid system has been missing and is one of the main goals of our paper. Pinning down the microscopic origin of the various relaxation and pure dephasing channels can provide a handle on engineering the various couplings to the environment to access the strong-coupling regime crucial for building highly efficient novel amplifiers for quantum microwave signals. 


The cavity-coupled DQD system studied here is in some ways analogous to a single-atom laser \cite{} where the cavity is formed by a superconducting transmission line resonator, the gain medium is formed by the voltage-biased DQD and pumping is achieved by current driven through the DQD. However, as we will show below, the intricate interplay between phonon and photon-assisted tunneling processes lead to a setting where electron and photon transport can no more be treated separately. From this perspective this system is rather similar to a Quantum Cascade Laser \cite{qcl,qclreview} where the active region is composed of a DQD structure. We first derive the Master equation for the reduced density matrix of the DQD-cavity system in the sequential tunneling regime with one extra electron injected through the leads. For this we start with a microscopic description of the coupling to the phonon bath that has recently been experimentally studied for InAs nanowire QDs \cite{weber10}. We derive analytical expressions for transmission, phase response, thresholds, lasing frequencies, photon number and steady state electron current. In particular, we show that the non-equilibrium optical susceptibility of the cavity-DQD amplifier features significant renormalization of the gain peak and bandwidth with respect to a bare current-driven DQD. Next we dissect the back action contributions of photon and phonon emission on the non-equilibrium current driven through the DQD. With recent progress in the measurement of photon correlation functions in superconducting circuits \cite{qs2,qs3,qs4}, it is now possible to analyze the statistics of emitted microwave photons. With this in mind, we compute the spectrum and the second order correlation function of photons leaving the cavity. Analytical results are benchmarked against exact-numerics and when applicable this is compared to existing experimental measurements. 

The paper is organized as follows. In Sec. \ref{modelDQD} we describe the system studied in recent experiments\cite{petta2014}. We derive the full Master equation for a voltage-biased DQD coupled to a single-mode cavity from a microscopic model. In Sec. \ref{PQTP} we derive an effective two-level model for the combined system that allows us to analyze it in the framework of a single-atom laser. We compute the laser threshold, frequency, spectrum, photon number and second order correlation function of the photon flux leaving the cavity. 
We subsequently compute the transmission and the phase response below the semiclassical laser threshold and compare to experimental measurements to establish the experimental regime of parameters.
In Sec. \ref{PPF} we calculate the non-equilibrium steady state current through the DQD and compare to experimental measurements. We devote Sec. \ref{validity} to the comparison of our analytical results with exact numerics and also, when applicable, with experimental measurements. In this section we also describe our findings on the emission spectrum and the second order correlation function of the photon flux leaving the cavity and its dependence on experimentally tunable parameters. The paper is summarized with an outlook in Sec. \ref{CON}. Details of some calculations are delegated to the appendices. Appendix \ref{appA} deals with the electron transport and treatment of electron leads as a bath for the cavity-coupled DQD system. In Appendix \ref{EOMapp}, we derive the equations of motion for our system and derive the various quantities of experimental and theoretical interest. 

\begin{figure}[t!]
	\centering
	\includegraphics[width=3.20in]{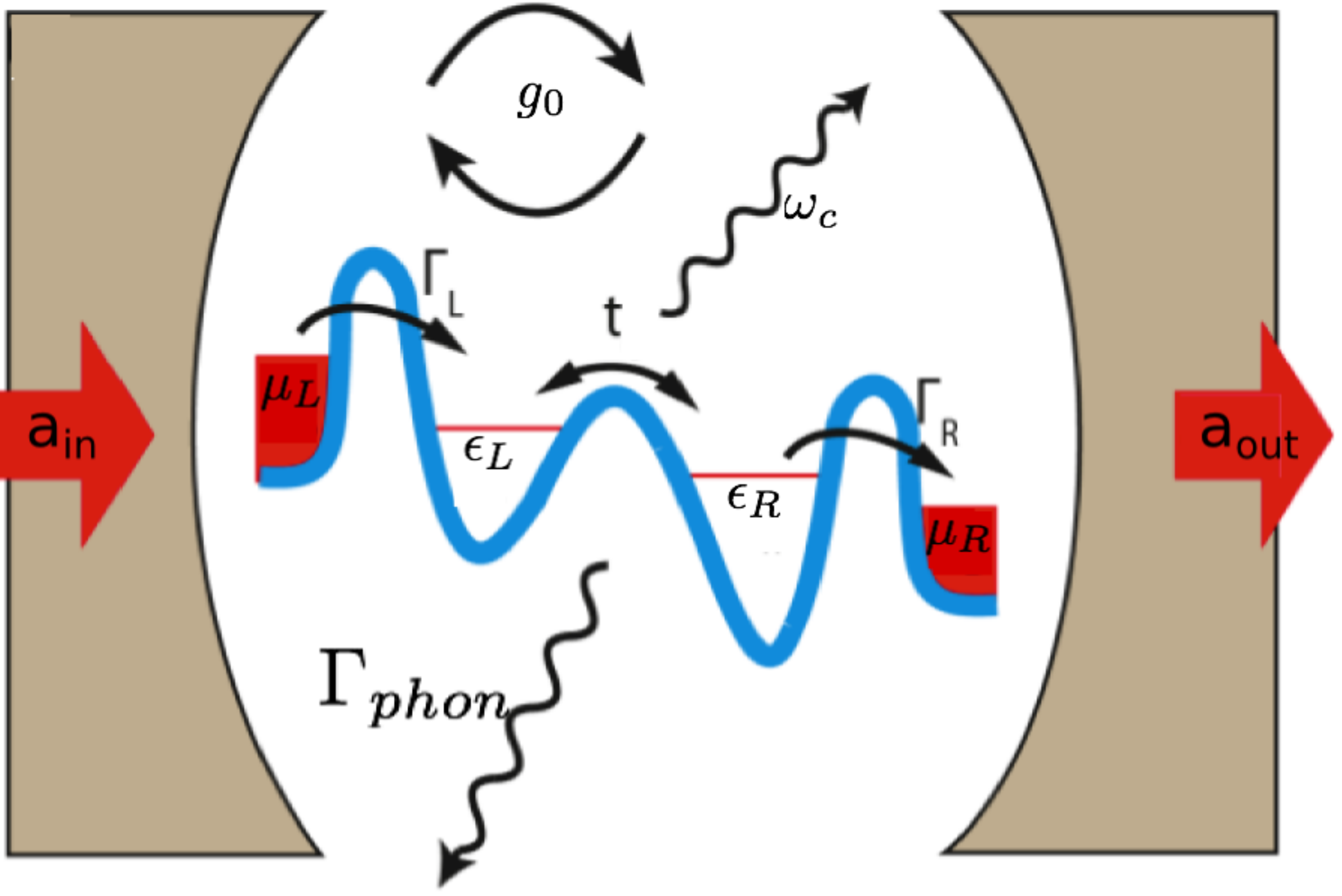}
         \includegraphics[width=3.17in]{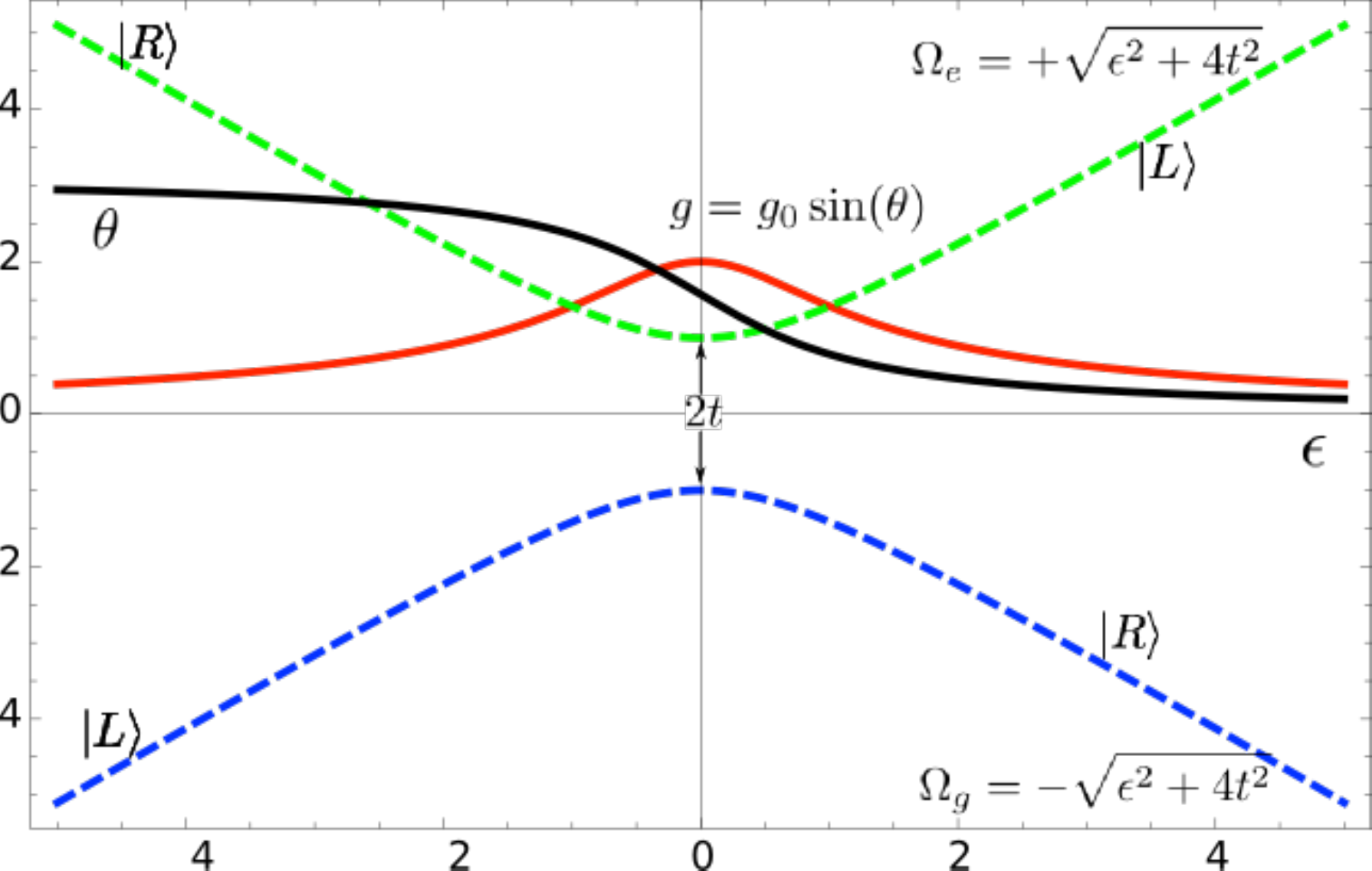}
	\caption{(upper panel) Schematic of the DQD cQED experiment\cite{petta2014}. The DQD is driven out of equilibrium by the application of a finite source-drain bias, which forces electrons to tunnel through the DQD. Due to electric dipole coupling between the trapped charge in the DQD and the electric field of the cavity, tunneling events in the DQD can influence the electromagnetic field inside of the microwave cavity and vice-versa. (lower panel) Energetics of the DQD and effective light-matter coupling $g$ (red) are shown as a function of detuning. Mixing angle $\theta$ as a function of detuning is seen in black (solid).}
	\label{fig:schem}
\end{figure}

\section{The Master equation of the DQD-Cavity system}
\label{modelDQD}

As schematically depicted in Fig. \ref{fig:schem} we consider a double quantum dot (DQD) that is connected to fermionic reservoirs (FR, chemical potentials $\mu_L$ and $\mu_R$, $V_{SD}= \mu_L - \mu_R$), dipole coupled (vacuum Rabi frequency $g$) to a transmission line resonator (R). The DQD is defined on an InAs nanowire (NW) and is coupled to phonon baths (PHON) of the substrate and the nanowire. We start with a microscopic description of the phonon-photon-DQD-FR 
\bea
\label{start_H}
H&=&H_{DQD}+H_{R}+H_{DQD-R}\nonumber\\&+&H_{PHON}+H_{DQD-PHON}\nonumber\\&+&H_{FR}+H_{FR-DQD}
\eea
Leaving out the parts involving leads, the remaining Hamiltonian can be written as 
\bea
\label{Hint}
H_{DQD,R,PH}&=&\int_{-\infty}^{\infty} dx \, \Psi^\dagger(x) \Bigg[  \frac{\hbar^2}{2m}\nabla^2 + V_{DQD}(x) \nonumber\\ &+& V_{PHOT}(x)+V_{PHON}(x) \Bigg] \Psi(x)
\eea
where
\bea
\Psi^\dagger(x)= e_{L}^{\dagger} \phi_L (x) + e_{R}^{\dagger} \phi_R (x) 
\eea
Here, $V_{DQD}(x)$ is an electrostatic potential describing the DQD and $\hat{\Psi}(x)$ is the field operator for DQD electrons. A realistic $V_{DQD}(x)$ can be obtained from a Hartree-Fock modeling of the DQD. Here we assume that only two energy levels of the DQD are relevant and use the variational basis $\phi_{L,R}(x)\sim e^{-\frac{(x\pm d/2)^2}{ 2 a^2}}$, where $d$ is the inter-dot distance  and $a$ denotes the dot size. Using orthogonality properties of the basis functions $\phi_L(x), \phi_R(x)$ we get,
\bea
H_{DQD}= \varepsilon_L e_L^\dagger e_L + \varepsilon_R e_R^\dagger e_R + t (e_L^\dagger e_R + e_R^\dagger e_L)
\eea
where $t$ is the tunnel coupling between the two dots given by
\bea
t\equiv \int_{-\infty}^{+\infty} dx \, V_{DQD}(x) \phi_L(x) \phi_R(x)
\eea
and $\varepsilon_i \equiv \int_{-\infty}^{+\infty} dx \, V_{DQD}(x) |\phi_i(x)|^2$ are the bare energy levels of the two QDs. The effective potential created on the quantum dot due to the cavity field is given by
\bea
\label{vphoton}
V^{L,R}_{photon}(x)= \alpha_{L,R} (a+a^{\dagger})
\eea
Plugging in Eq. \ref{vphoton} into Eq. \ref{Hint} gives the light-matter coupling within the constant interaction model
\bea
H_{DQD-R} =g_0 \tau_z (a+a^{\dagger})
\eea
where $g_0$ is the coupling strength and $\tau_z \equiv \ket{L}\bra{L} - \ket{R}\bra{R}$.  

We next study the effective potential created by phonons. There two important phonon contributions\cite{weber10}, those of the substrate (deformation potential and piezoelectric) and the nanowire. This results in a total phonon spectral function
\bea
\label{jtot}
J(\omega) = J_{B,piezo}(\omega)+J_{B,dp}(\omega)+J_{NW}(\omega)
\eea
where we have
\bea
\label{jdef}
J_{B,dp}(\omega)=j_{dp}  \left(\frac{\omega}{\omega_0}\right)^3 e^{-\frac{\omega^2 }{ \omega_{D}^2}}\left(1- \mathrm{sin}\left(\frac{\omega }{\omega_{DD}}\right)\right)\\
J_{B,piezo}(\omega)=j_{piezo}  \left(\frac{\omega}{\omega_0} \right)e^{-\frac{\omega^2 }{ \omega_{D}^2}}\left(1- \mathrm{sin}\left(\frac{\omega }{\omega_{DD}}\right)\right) .
\label{jpiezo}
\eea
and $J_{NW}(\omega)$ can be obtained from modeling with experimentally extracted parameters\cite{weber10}. In Eq. \ref{jdef} and Eq. \ref{jpiezo}, $\omega_D=\frac{\sqrt{2}c_s}{a_z}$ and $\omega_{DD}=\frac{c_s}{d}$ and $\omega_0$ is a scaling parameter introduced for convenience which we choose to be $32.8\mu eV$. Here $c_s$ is the speed of phonon propagation ($c_s\sim 11000$ m/s) and $d$ is the distance between quantum dots ($d\sim$ 120 nm for our structure), each with an axial confinement (dot size) given by $a_z\sim 40-50$ nm and the radius of the nanowire is estimated to be around $r\sim50$ nm. 

The coupling strength constants $j_{piezo}$ and $j_{dp}$ are hard to independently estimate with the available experimental data\cite{petta2014,weber10}. However, for the purpose of this paper we choose $j_{piezo} \approx 5.96$ $\mu$eV and $j_{dp} \approx 0$ to fit the available experimental data. We do not find significant change in the features if both are taken to be non-zero except in the asymptotic behavior which we elaborate on in later sections. 

After incorporating the effect of cavity photons, phonons and the DQD potential, one is left with having to treat the coupling with electron leads at finite source-drain bias.  For the treatment of the lead electrons, the reader is referred to Appendix \ref{appA}. 

After integrating out the lead electrons and phonons we arrive at the following master equation for the DQD-cavity system with a coherent microwave drive of frequency $\omega_d$ incident from the left waveguide: 
\bea\label{RWAMaster}
\dot{\rho} &=& - i [H,\rho] + \gamma_\downarrow \mathcal{D}(\sigma_-) +\gamma_\uparrow \mathcal{D}(\sigma_+) +\gamma_\phi\mathcal{D}(\sigma_z) +\kappa D(a) \nonumber\\&+& \Gamma_L  \cos^2 \left(\frac{\theta}{2}\right)\mathcal{D}(\ket{e}\bra{0})+\Gamma_L  \sin^2 \left(\frac{\theta}{2}\right)\mathcal{D}(\ket{g}\bra{0})\nonumber\\
& &+  \Gamma_R  \sin^2 \left(\frac{\theta}{2}\right)\mathcal{D}(\ket{0}\bra{e})+\Gamma_R  \cos^2 \left(\frac{\theta}{2}\right)\mathcal{D}(\ket{0}\bra{g})\nonumber\\
H &=&\frac{\Omega }{2}\sigma_z + \omega_c a^\dagger a + g (\sigma_+ a + \sigma_- a^\dagger) \nonumber\\&+& i\sqrt{\frac{\kappa}{2} } E \cos (\omega_d t) \, (a^\dagger - a) ,
\label{GeneralMaster1}
\eea
The system Hamiltonian (\ref{GeneralMaster1}) takes the form of a coherently driven Jaynes-Cummings model for a two level system with transition frequency $\Omega=\sqrt{\epsilon^2+4 t^2}$. Here $\sigma_{z,\pm}$ are pseudo spin operators for the diagonal DQD levels $e$ and $g$
\bea
\ket{e}&=&\cos\left(\frac{\theta}{2}\right)\ket{L}+\sin\left(\frac{\theta}{2}\right)\ket{R}\\ \ket{g}&=&-\sin\left(\frac{\theta}{2}\right)\ket{L}+\cos\left(\frac{\theta}{2}\right)\ket{R}
\eea 
with $\theta=\arctan(2t/\epsilon)$ and $g=g_0 \sin (\theta)$ is the $\epsilon$-dependent effective vacuum Rabi frequency (see Fig. \ref{fig:schem}). The last term in the Hamiltonian Eq. \ref{GeneralMaster1} is the coherent microwave tone of amplitude $E$ incident from the left waveguide (we assume the coupling to the left and right waveguides are symmetric, $\kappa_L = \kappa_R = \kappa/2$). 

Turning to the dissipative terms in (\ref{GeneralMaster1}) described by the Lindblad dissipators $\mathcal{D}(C)= C \rho C^\dagger - [C^\dagger C, \rho]/2$, $\kappa=\kappa_L + \kappa_R$ is the total cavity loss rate to the left and right waveguides and $\Gamma_{L,R}$ are the tunneling from the source reservoir (L) into the left dot and from the right dot into the drain reservoir (denoted by R). We consider a situation (See Fig. \ref{fig:schem}) where $V_{SD}=\mu_L - \mu_R >0$ and that for any left-right detuning $\epsilon \equiv \varepsilon_L - \varepsilon_R$ the levels are always far from the respective Fermi levels. Here, we assume that we operate within the finite bias triangle corresponding to a transport cycle with one additional electron sequentially tunneling through the DQD from left to right. While $\ket{L}$ and $\ket{R}$ denote single particle levels, we use it here in this regime also to denote the many-body energy levels with one additional electron on the left dot and the right dot, respectively. Additionally $\ket{0}$ denotes the ``empty'' dot i.e. the dot in its original configuration before the injection of an electron. Finally, $\gamma_\downarrow$, $\gamma_\uparrow$, $\gamma_\phi$ are the phonon relaxation, pumping and pure dephasing contributions that follows from Eq. \ref{jtot} given by 
\bea
\label{up}
\gamma_\uparrow (\epsilon) &=&\sin^2(\theta) \bar{n}(\Omega,T) J(\Omega) \\
\label{down}
\gamma_\downarrow (\epsilon) &=&\sin^2(\theta)  (1 + \bar{n}(\Omega,T)) J(\Omega) \\
\label{g_phi}
\gamma_\phi (\epsilon) &=& \cos^2(\theta) (1+2 \bar{n}(0,T))  J(0)
\eea
We note that these rates are strong functions of the left-right detuning $\epsilon$ which will have important ramifications for the results obtained. Here $\bar{n}(\omega,T)=\frac{1}{e^{\omega/T}-1}$  denotes the mean phonon occupation at frequency $\omega$ and temperature $T$. We assume that $T$ is finite but small with respect to all energy scales of the Hamiltonian. Because $J \sim \omega^s$ with $s>0$, the pure dephasing term $\gamma_\phi$ can be neglected (see also Ref.~\onlinecite{aguado, tureci_nanowire}). In what follows, all numerical data presented and compared to the analytic and experimental quantities are computed through the exact numerical solution (with a cavity photon number cutoff) of  Eq. \ref{GeneralMaster1}.

\section{Analogy with single atom laser and beyond}
\label{PQTP}

\begin{figure}[t!]
	\centering
	\includegraphics[]{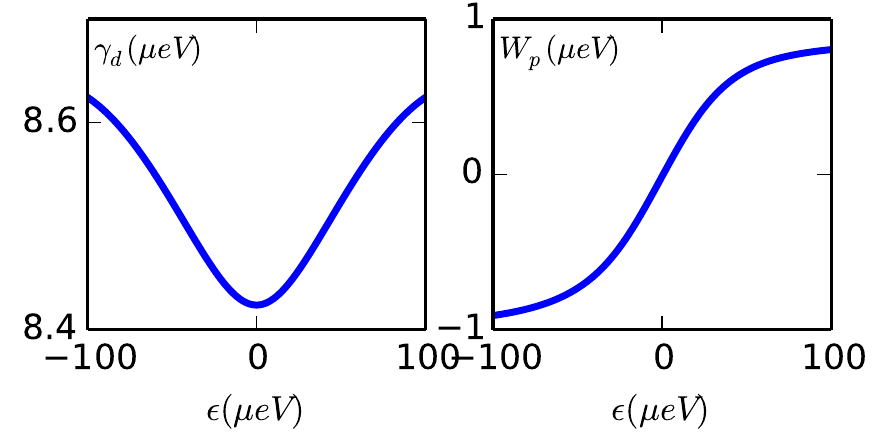}
	\caption{(Left panel) $\gamma_{d}$ (Eq. \ref{gamma_d}) and (Right Panel) $W_p$ (Eq. \ref{inv}) as a function of detuning $\epsilon$.}
	\label{fig:WP0}
\end{figure}

The Master equation derived in the previous section is not convenient to gain an analytic insight into the physics of the DQD-cavity amplifier. Progress can be made by integrating out the empty dot state and setting up an analogy to the standard theory of the single-atom laser/amplifier \cite{cmb}, 
\bea
\label{sal1}
\dot{\rho} &=& - i [H,\rho] \nonumber\\&+& \Gamma_\downarrow \mathcal{D}(\sigma_-)+ \Gamma_\uparrow \mathcal{D}(\sigma_+) +\Gamma_\phi \mathcal{D}(\sigma_z) + \kappa D(a),\\
\label{sal2}
\eea
with effective pumping, relaxation and pure dephasing rates given by\footnote{These effective rates are plotted in Figure \ref{fig:WP} (appendix) as a function of detuning.}
\bea
\Gamma_\uparrow = \gamma(1+W_p)/2\\
\Gamma_\downarrow = \gamma(1-W_p)/2\\
\Gamma_\phi =( \gamma_d - \gamma/2)/2
\eea
where (also see Fig. \ref{fig:WP0})
\bea
\label{gamma_gamma}
\gamma&=&\gamma_{\uparrow}+\gamma_{\downarrow}+\Gamma_{R}\left[\cos^{4}\left(\frac{\theta}{2}\right)+\sin^{4}\left(\frac{\theta}{2}\right)\right]\nonumber\\&-&\frac{\cos\theta}{1+\frac{\Gamma_{R}}{2\Gamma_{L}}}\frac{\Gamma_{R}}{2\Gamma_{L}}\left[\gamma_{\uparrow}-\gamma_{\downarrow}+\Gamma_{R}\cos\theta\right]\\
W_p&\equiv&\frac{1}{\gamma}\frac{\gamma_{\uparrow}-\gamma_{\downarrow}+\Gamma_{R}\cos\theta}{1+\Gamma_{R}/2\Gamma_L} 
\label{inv}\\
\gamma_{d} &=& 2\gamma_\phi+(\gamma_\uparrow+\gamma_\downarrow+\Gamma_{R})/2
\label{gamma_d}
\eea
This master equation describes a single-atom laser with active levels $e$ and $g$ and effective pump and dissipation terms that are complicated functions of the inter-dot detuning $\epsilon$, through the dependences of the mixing angle $\theta(\epsilon)$ and the phonon dissipation rates $\gamma_\downarrow (\epsilon)$, $\gamma_\uparrow(\epsilon)$, $\gamma_\phi(\epsilon)$. We note that the steady-state solution $\rho_{ss}$ of this Master equation contains full information about the photonic as well as the electronic transport observables. In this section we focus the photonic aspects and in particular, the transmission, phase response, photon number, emission spectrum and the photon statistics. The reader is referred to Appendix \ref{EOMapp} for the details of calculations leading to the transmission and the phase response.

Turning the drive off ($E=0$) we obtain the following equations of motion in the rotating frame with respect to $\omega_l$, where $\omega_l$ is an emergent frequency scale to be determined below:
\bea
\langle \dot{a} \rangle &=& - i \Delta_{cl} \langle a \rangle - \frac{\kappa}{2}\langle a \rangle - i g \langle\sigma_- \rangle 
\label{eqef},
\\ \label{eqe1}
\langle \dot{\sigma}_- \rangle &=& -(i \Delta_{ql} + \gamma_{d})\langle \sigma_- \rangle+ i g \langle a \sigma_z\rangle ,\\
\label{eqe2}
\langle \dot{\sigma}_z \rangle&=& \gamma (W_p - \langle \sigma_z\rangle) +2 i g (\langle a^\dagger \sigma_-\rangle -\langle a \sigma_+\rangle)
\eea
Here, $\Delta_{cl}=\omega_c-\omega_l$,  $\Delta_{ql}=\Omega-\omega_l$, $W_p(\epsilon)$ is the effective incoherent pump power measured in terms of the unsaturated inversion it creates, $\gamma_{d} (\epsilon)$ is the dipole dissipation rate and $\gamma (\epsilon)$ is the population relaxation rate (see Eqs. (\ref{gamma_gamma}) - (\ref{gamma_d})). We show the strong $\epsilon$-dependence of the pump rate $W_p$ and the dipole dissipation rate $\gamma_{d}$ in Fig.~\ref{fig:WP0}.


These equations can be solved in the steady-state for $\langle a \rangle_{ss}$, $\langle\sigma_- \rangle_{ss}$, $\langle\sigma_z \rangle_{ss}$ by the semiclassical factorization $\langle a O \rangle \approx \langle a \rangle \langle O \rangle$ yielding equations that are equivalent to Maxwell-Bloch equations. In this approximation equations \ref{eqef} to \ref{eqe2} reduce to
\bea \label{a_no_d}
\langle a \rangle_{ss}\left( 1 - \frac{1}{\frac{\kappa}{2} + i \Delta_{cl}}\cdot \frac{g^2}{i \Delta_{ql} + \gamma_{d}} \cdot\frac{W_p}{1 + G |\langle a \rangle_{ss}|^2} \right) = 0 \nonumber
\eea
which is an equation for $|\langle a \rangle_{ss}|$ and $\omega_l$ leaving the phase of the field undetermined. Here $G=\frac{4 g^2 \gamma_d}{\gamma (\gamma_d^2+\Delta_{ql}^2)}$. These can be solved and yield the solution 
\bea\nonumber \label{PHOT}
\langle a^\dagger a \rangle_{ss} =  | \langle a \rangle_{ss} |^2 &=&  0 ,  \,\,\,\,\qquad\qquad W_p <W_{th} \\ 
&  = &\frac{1}{G} (\frac{W_p}{W_{th}}-1),  W_p >W_{th}
\label{phnum}
\eea
Hence, this semiclassical solution predicts lasing above the threshold pump power $W_p = W_{th} = \frac{\kappa\left(\gamma_{d}^{2}+\Delta_{ql}^{2}\right)}{2g^{2}\gamma_{d}}$. The same equation provides the line-pulling formula for the lasing frequency 
\bea
\omega_l = \omega_c \frac{\gamma_d}{\gamma_d + \kappa/2} + \Omega \frac{\kappa/2}{\kappa/2 + \gamma_d} .
\eea
The steady state inversion is clamped to its threshold value above the threshold
\bea
\label{sigz}
\langle \sigma_z \rangle_{ss}=
W_p, \nonumber& W_p <W_{th}\\= W_{th}, & W_p >W_{th}
\label{lpf}
\eea

\begin{figure}[t!]
	\centering
	\includegraphics[]{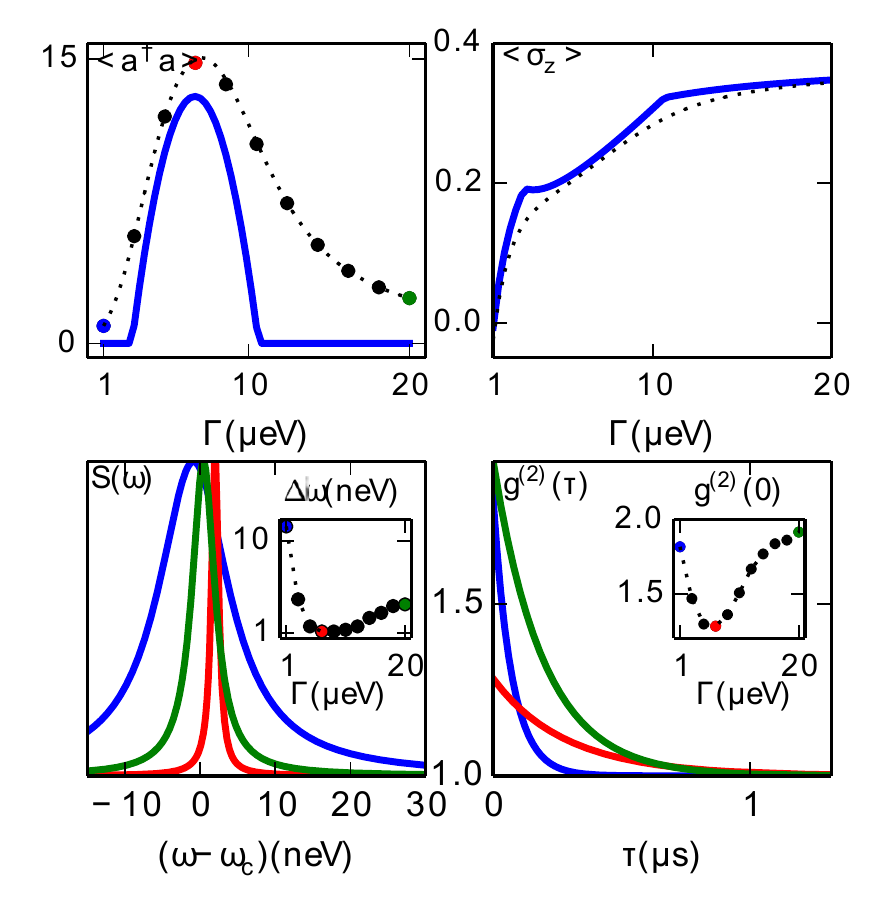}
	\caption{Power spectrum and $g^{(2)}(\tau)$. The values are the same as in Table I except where mentioned. 
		(Left panel) Plot showing spectrum $S(\omega) = \int_{- \infty}^{ \infty} e ^{i \omega t} \langle a^{\dagger} (t) a (0) \rangle dt$ with varying dot-lead tunneling rate. (Right) Plot of $g^{(2)}(\tau)=\frac{\left\langle a^{\dagger}a^{\dagger}(\tau)a(\tau)a\right\rangle }{\left\langle a^{\dagger}a\right\rangle ^{2}}$ for different dot-lead tunneling rates. We consider the symmetric case $\Gamma_L = \Gamma_R =\Gamma$.}
	\label{fig:sg2}
\end{figure}

It's important to note that the semiclassical solutions have a limited validity for a single atom laser. This stems from the fact that the gain medium is not a collective system described by a large spin but a single spin for which quantum fluctuations cannot be ignored. To illustrate this we plot in Fig. \ref{fig:sg2} the semiclassical photon number versus the exact photon number obtained by the numerical solution of the Master equation (Eq. \ref{GeneralMaster1}) as well as the inversion. Here we ramp up the pump power by biasing the gates between the left dot and the source reservoir (i.e. $\Gamma_L$). 

We note the existence of a second threshold above which lasing turns off. This stems from the quantum nature of the gain medium. The same bath that via a complex interplay of phononic and fermionic baths gives rise to the pumping term ($W_p$), also gives rise to the quenching of the DQD dipole (via $\gamma_d$). While $W_p$ saturates (as e.g. $\Gamma_L$ is ramped up) $\gamma_d$ monotonically increases. Above some critical tunneling rate lasing is quenched, signaled by a sharp semiclassical second threshold visible in Fig. \ref{fig:sg2}. It's important to note that the laser threshold can only be reached in the strong coupling regime $g^2 > \kappa \gamma_d$. 

A further indication that what we have here is not a conventional laser can be seen from the emission spectrum and in particular the photon statistics. In Fig. \ref{fig:sg2} we show the emission spectrum $S(\omega)$ and the second order correlation function $g^{(2)}(\tau)$ of photons emitted from the cavity: 
\bea
S(\omega) = \int_{- \infty}^{ \infty} e ^{i \omega t} \langle a^{\dagger} (t) a (0) \rangle dt
\eea
and
\bea
g^{(2)}(\tau)=\frac{\left\langle a^{\dagger}a^{\dagger}(\tau)a(\tau)a\right\rangle }{\left\langle a^{\dagger}a\right\rangle ^{2}}
\eea
We note that the broad emission spectrum of the DQD-cavity system below the semiclassical threshold narrows by an order of magnitude as we approach maximum cavity emission, and then broadens again. The behavior above the threshold is consistent with the Shawlow-Townes linewidth $\Delta\omega \propto 1/ \langle a^\dagger a \rangle_{ss}$ \cite{}.  We also note that there is a strong frequency renormalization of the spectral peak from the cavity frequency below the threshold to the laser frequency given by the line-pulling formula Eq. \ref{sigz}. Turning to the photon statistics, we see in the right panel that below the threshold we have effectively a thermalized photon gas in the cavity ($g^{(2)}(0) \sim 2$) while above the threshold the distribution $g^{(2)}(0)$ moves away from its thermal value but remains always above 1. For larger $\Gamma$, the distribution asymptotically approaches 2 again.

\begin{figure}[t!]
	\centering
	\includegraphics[]{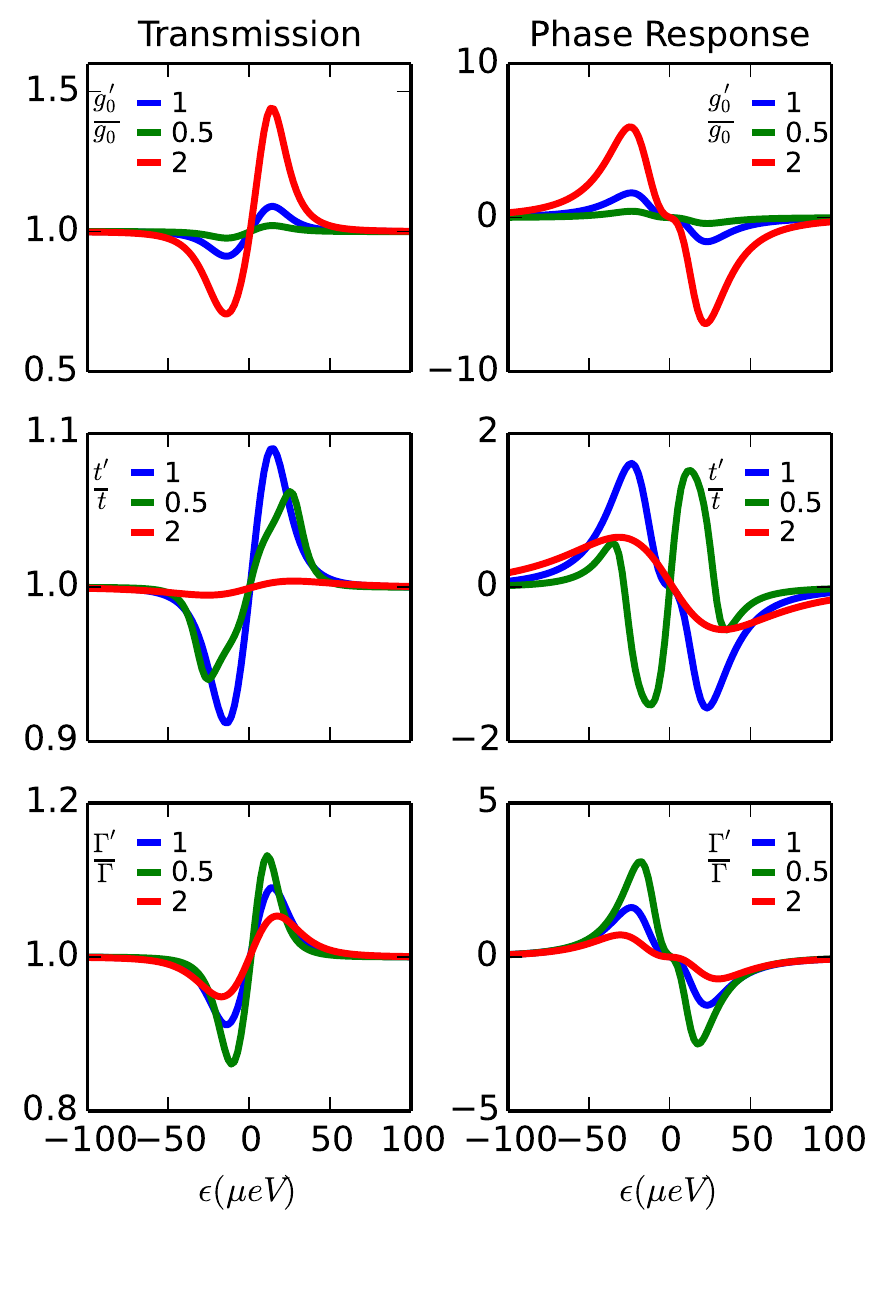}
	\caption{ 	Plot showing analytical results for transmission (left column) and phase response (right column), ie, argument in degrees of  Eq. \ref{TRANS} versus detuning. From top to bottom panel are the behaviour as one changes (i) Light matter coupling $g_0$. (ii) Interdot tunneling $t$ and  (iii) dot lead tunneling rates $\Gamma_L,\Gamma_R$. The analytical plot are generated using Eq \ref{TRANS}}
	\label{fig:tranphase}
\end{figure}

We now consider the transmission   $A =  \frac{a \sqrt{2\kappa } }{E }$  by turning on $E$ while remaining in the linear response regime. This can be obtained by the non-linear steady-state susceptibility of the amplifier (See Appendix) 
\bea\label{TRANS}
A (\epsilon)= \frac{i\kappa/2}{i\kappa/2+ \Sigma(\epsilon)}
\eea
\bea
Im[\Sigma]&=&\frac{-g^2 W_{p} \gamma_d}{\Delta_{qd}^2+\gamma_d^2}\\
Re[\Sigma]&=&-\frac{g^2 W_{p} \Delta_{qd}}{\Delta_{qd}^2+\gamma_d^2}
\eea
In what follows, instead of analyzing the $\omega_d$ dependence of the transmission amplitude, we will fix the incident microwave signal to the cavity frequency $\omega_d = \omega_c$ and analyze the dependence on the two-level system transition frequency via the tuning of $\epsilon$. The self-energy correction emerges due to the coupling of the DQD electron to phonon and electron baths. In Fig. \ref{fig:tranphase} we show the transmission and phase response as a function of detuning. In particular, we show the trend as one increases the lead-dot tunneling rate ($\Gamma'$), light matter coupling ($g'$) and the interdot tunneling ($t'$). As reference ($g$,$t$,$\Gamma$) we take the experimental values for data analyzed in section \ref{validity}.  

The most distinctive feature of the transmission amplitude is the gain peak on the positive side of detuning $\epsilon>0$. There is a significant renormalization of the position of the gain peak with respect to a bare current-driven DQD which would be at resonance, $\omega_d = \Omega$ or $\epsilon_{opt} =  \sqrt{\omega_d^2 - 4t^2}  $. Assuming the relaxation rate arising due to phonons is small compared to the dot-lead tunneling rate (as is the case for data presented in Section. \ref{validity}), we arrive at the following expression for the optimal $\epsilon_{opt}$ at which maximal gain results: 
\bea
\epsilon_{opt} \sim\frac{2 \sqrt{2} t \sqrt{-16 t \omega _c+4 \omega _c^2+\Gamma ^2+16 t^2}}{\sqrt{-224 t \omega _c+44 \omega _c^2+11 \Gamma ^2+272 t^2}}
\label{epscritical}
\eea
We note that this optimal value is the result of the saturation of the effective pump rate (Fig \ref{fig:WP0}) and the quenching of the light-matter interaction strength $g$ with increasing $\epsilon$ (Fig \ref{fig:schem}). The above expression is valid only when the observed maximal gain is at relatively small detuning. For the realistic parameters of the experiment, the above formula for optimal detuning seems to be the point where the peak is theoretically predicted. 

\section{Non-equilibrium electron current}
\label{PPF}

In this section we analyze the non-equilibrium steady state current below and above the laser threshold. In the presence of non-zero phonon and photon coupling, a rather compact analytic expression can be obtained in the small $\Gamma/t$ limit 
\bea\label{CUR}
\label{small-gamma}
I^{\frac{\Gamma}{t}<1} &=& \frac{\Gamma_R}{2(1 + \Gamma_R/2\Gamma_L)} (1 - \langle \sigma_z \rangle_{ss} \cos\theta) + \Gamma_R \sin\theta\,\Re (\langle \sigma_- \rangle_{ss}) \nonumber\\&=&  \frac{\Gamma_R}{2(1 + \Gamma_R/2\Gamma_L)} (1 - \langle \sigma_z \rangle_{ss} \cos\theta) ,
\eea
This expression is valid both below and above threshold. Here we can insert $\langle \sigma_z \rangle$ from Eq. \ref{sigz} and $\langle \sigma_- \rangle = 0$. Eq. \ref{CUR} has a complex combination of phonon and photon contribution that can be isolated. 


In the limit of very small phonon and photon coupling the current is mediated only by coherent inter-dot tunneling and is given as a simplified limit of Eq. \ref{small-gamma} by
\bea
\label{loren_i}
I^{\mbox{no phonon, no photon}}=\frac{\text{\ensuremath{\Gamma_{L}}}\text{\ensuremath{\Gamma_{R}}}t^{2}}{\text{\ensuremath{\Gamma_{L}}}\epsilon^{2}+t^{2}(2\text{\ensuremath{\Gamma_{L}}}+\text{\ensuremath{\Gamma_{R}}})}
\eea

Below the lasing threshold, we have only phonon contribution (in addition to the coherent inter-dot tunneling contribution) and this gives us, the following
\bea
\label{phonexp}
I^{\mbox{PHON}}=\frac{\Gamma_{L}\Gamma_{R}t^{2}}{\Gamma_{L}\epsilon^{2}+t^{2}(2\Gamma_{L}+\Gamma_{R})}\qquad\qquad\qquad\quad\nonumber\\+\gamma_{\downarrow}\frac{\Gamma_{L}^{2}\epsilon\left[2t^{2}\left(\sqrt{4t^{2}+\epsilon^{2}}+2\epsilon\right)
+\epsilon^{2}\left(\sqrt{4t^{2}+\epsilon^{2}}+\epsilon\right)\right]}{2\left(\text{\ensuremath{\Gamma_{L}}}\epsilon^{2}+t^{2}(2\text{\ensuremath{\Gamma_{L}}}+\text{\ensuremath{\Gamma_{R}}})\right)^{2}}
\nonumber\\+O(\gamma_{\downarrow}^2)\qquad\qquad\qquad
\eea
One important consequence of including the phonon contribution is the induced asymmetry in the current-detuning dependence as opposed to a symmetric lorentzian obtained (Eq. \ref{loren_i}) when phonons are neglected.

\begin{figure}[t!]
	\centering
	\includegraphics[]{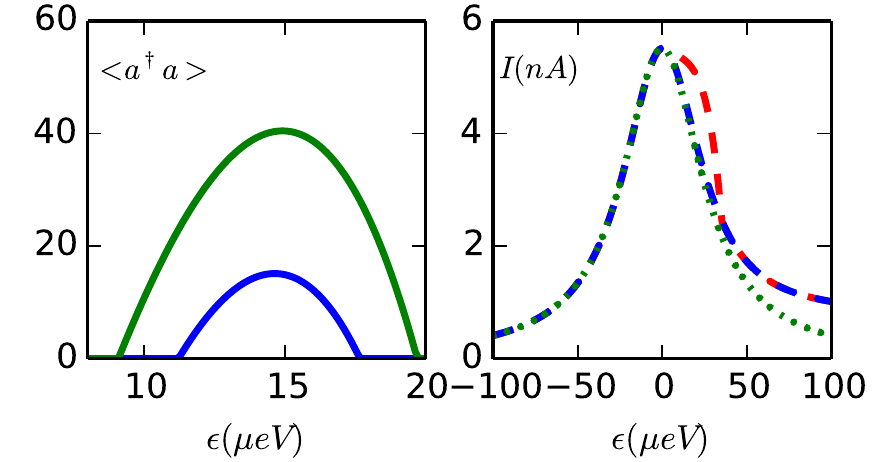}
	\caption{(Left) Plot showing photon number versus detuning with and without phonons. In particular, one clearly notices the phonon reduced photon emission. Green is without phonons and blue is with phonons. We used $g_0'=5g_0$. (Right) Plot showing current (analytical) versus detuning. The three plots of nonequilibrium current here are (green) without phonons and photons, (blue) with only phonons and (red) with both phonons and photons We used a light matter coupling $g_0'=10g_0$. The analytical plots are generated using Eq \ref{PHOT} and \ref{CUR}}
	\label{fig:flux}
\end{figure}

 In Fig. \ref{fig:flux}, we show the phonon and photon mediated current contributions to the total non-equilibrium steady state current as a function of detuning. Close to the peak current on the positive detuning side, the photon-mediated component gives rise to an additional current due to the stimulated emission being maximal in that range. This additional photon-induced current on the positive detuning side (see right panel of Fig. \ref{fig:flux}) occurs in a detuning window where the system is above the lasing threshold, 
\bea
\epsilon:\, W_{P}(\epsilon)>W_{th}(\epsilon)
\eea

It is worthwhile to analyze the asymptotic behaviour of the DC current Eq. \ref{small-gamma}. For large detuning $\epsilon$ the threshold is pushed further such that $W_{P}(\epsilon)<W_{th}(\epsilon)$. This means that one is left only with the coherent tunneling current and the phonon induced current.

The tail on the positive detuning side is dominated by phonon-induced processes. We note that for $\epsilon \rightarrow \infty$, the current behaves as
\bea
\label{farcurr}
I_{|\epsilon|\rightarrow \infty}=\frac{\Gamma_R t^2}{\epsilon^2}+\gamma_{\downarrow}
\eea
Therefore, the large detuning tail of the current (Eq. \ref{farcurr}) is a competition between coherent tunneling, deformation and piezo phonon contribution, ie, $\epsilon^{-2}$, $\epsilon\, e^{-\epsilon^2/\omega_D^2}$ and $ \epsilon^{-1} e^{-\epsilon^2/\omega_D^2}$ respectively.


%
%
%

\section{Experiments measurements and comparison to theoretical predictions}
\label{validity}

\begin{figure}[h!]
\includegraphics[]{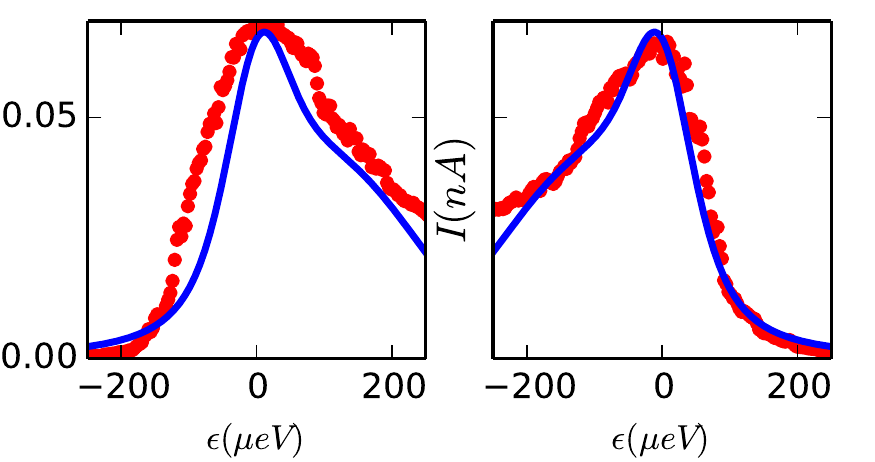}
\caption{
Comparison of our non-equilibrium current analytical expression (Eq. \ref{small-gamma}) through a system with small DQD-leads coupling for negative (left) and positive (right) source drain bias with the experimental data. The best fit was for $t=29$ $\mu$eV, $c_{piezo}' = 1.7\mu$eV, $\Gamma_R = 0.77 \mu$eV, $\Gamma_L=1.8 \mu$eV. Experiment is in blue cross and theory is red solid lines. The analytical fits are generated using Eq \ref{CUR}.}
\label{fig:Small1}
\end{figure}

\begin{figure}[h!]
\includegraphics[]{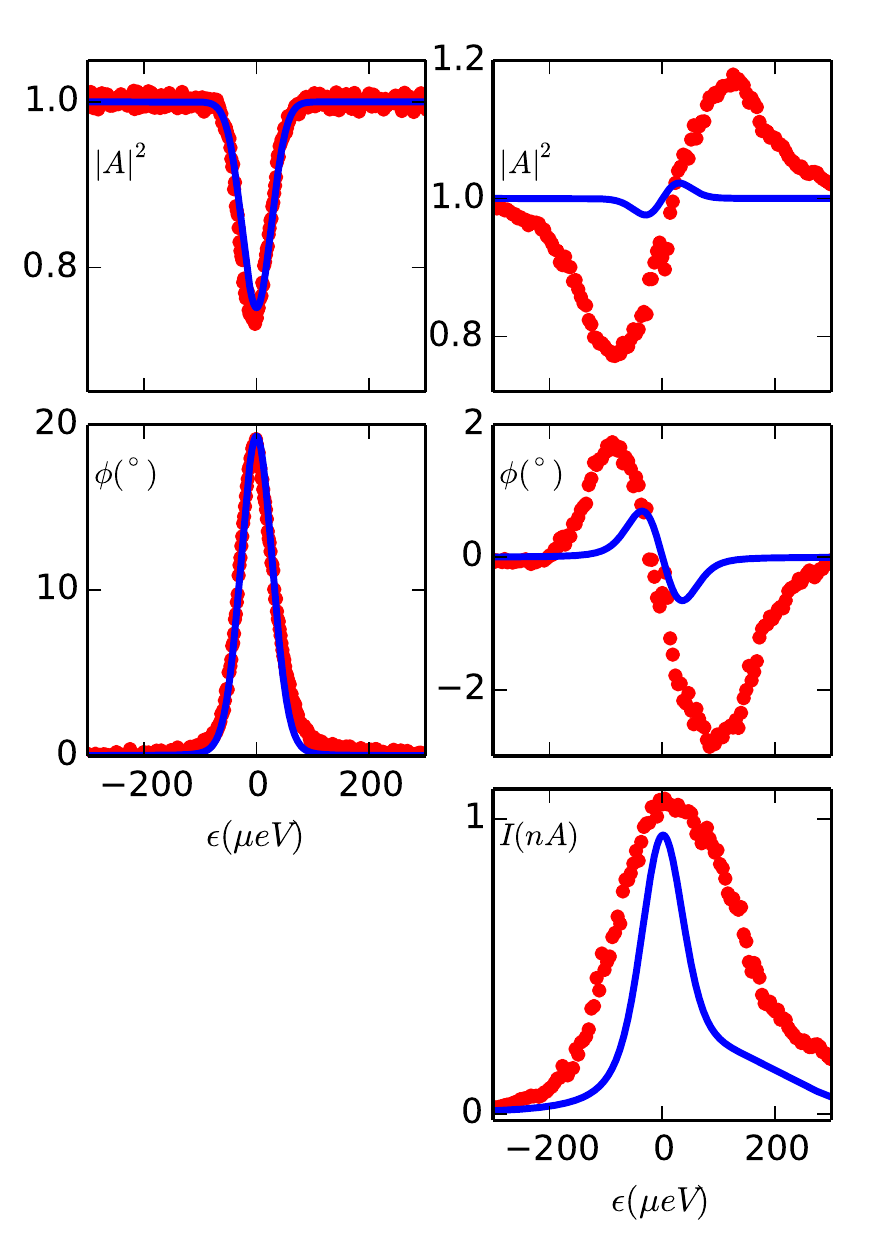}
\caption{Theoretical vs experimental fit.  The first column represents zero bias data. The second column represents finite bias data. The first row represents transmission, the second phase response and the last current (obviously, there is no current in the zero bias regime). The analytical fits are generated using Eq \ref{TRANS} and \ref{CUR}.}
\label{fig:RenTran}
\end{figure}

In this section, we first present experimental data (Fig. \ref{fig:Small1} and Fig. \ref{fig:RenTran}) and compare to predictions of the theoretical models discussed in the previous sections. Here we take into account charge noise of magnitude $\sigma \approx 25 \mu eV$ \cite{petta2014} by averaging any theoretical result $F(\epsilon)$ by $\bar{F}(\epsilon)=\int_{-\infty}^{\infty} d\tilde{\epsilon} \, e^{-\frac{\left(\epsilon-\tilde{\epsilon}\right)^{2}}{2\sigma^{2}}}F(\tilde{\epsilon})$. We will also compare our analytical results with exact numerical solution of the full Master equation (Eq. \ref{RWAMaster}). We analyze (i) Transmission versus detuning $\epsilon$, (ii) Phase response versus detuning, (iii) Current versus detuning and (iv) Photon number versus detuning. We find that the analytical results show perfect agreement with the numerically exact treatment of the master equation for the transmission, phase response and DC current. 
A more detailed analysis of the analytical results for transmission and phase response are presented in Fig. \ref{fig:tranphase}. In particular, one clearly sees that height (gain in signal) of transmission curve is non-monotonically dependent on both the inter-dot tunneling and the dot-lead tunneling rate. However, increasing the light-matter coupling enhances the transmission signal.

As expected the photon flux in the semiclassical treatment above the semiclassical threshold quantitative differs from exact numerical results. Semiclassical equations of motion are not expected to be valid near the threshold.

Fig. \ref{fig:flux} (left) shows the behaviour of photon flux for two different scenarios which include and exclude phonons. One can clearly see the phonon suppressed photon flux here. The right part of Fig. \ref{fig:flux} shows the DC current (i) without phonons and photons (ii) with only phonons and (iii) with both phonons and photons. The photon and phonon assisted nonequilibrium DC electron transport is clearly visible in Fig. \ref{fig:flux}. 

\bigskip
\captionof{table}{Parameter values from Ref. \onlinecite{petta2014}}  \label{tab:title}
\begin{tabularx}{0.5\textwidth}{l  X }
\hline
Cavity and driving frequency $\omega_c$&  $32.5$ $\mu$eV \\ 
Cavity loss rate $\kappa$ & $0.0082$ $\mu$eV\\
Light-matter coupling $g_0$& $0.0662$ $\mu$eV \\
Elastic tunneling $t$ & $16.4$ $\mu$eV\\
Drain tunneling rate $\Gamma_R$ & $16.56$ $\mu$eV \\ 
Source tunneling rate $\Gamma_L$ & $16.56$ $\mu$eV \\
Phonon cutoff frequency $\omega_{cutoff}$ & $256$ $\mu$eV\\
$\omega_D$ & $60$ $\mu$eV \\
Scaling frequency $\omega_0$ & $32.8$ $\mu$eV\\ 
$j_{piezo}$ (unknown)  & $5.96$ $\mu$eV \\ 
$j_{def}$ (unknown) & $0.0$ $\mu$eV \\ 
Speed of sound in $SiN$ $c_s$ & $11000$ m/s\\
Size of an individual quantum dot $a$ & $50$ nm\\
Temperature $T$ & $8$ mK \\
Gaussian noise $\sigma$& $25$ $\mu$eV\\
\end{tabularx}\par
\vspace{10 mm}
We next compare to existing experimental measurements of DC electron transport, transmission\cite{petta2014} and phase response. In Fig. \ref{fig:Small1}, we fit the analytical expression for current with the experimental data for the scenario in which the dot-lead tunneling rates and hence the DC electron current are small ($\sim 0.07 nA$). One clearly sees that the asymmetry in DC current (as a function of detuning) is explained by the presence of phonons. This regime (Fig. \ref{fig:Small1}), ie, weak transport regime is explained well with our model which assumes a DQD weakly coupled to leads. In Fig. \ref{fig:RenTran}, we see that there is a sizable discrepancy between the theoretical data and experiments and we believe that this can be attributed to the strong coupling of the dot to the leads which invalidates the perturbative treatment of the leads underlying our Master equation.

%
%

%
%
%
%

\section{Conclusion}
\label{CON}
We provided a first-principles derivation of the Master equation for a DQD coupled to a single cavity mode, considering systematically both the presence of phonons and the effect of leads at finite voltage bias. We derived an effective model of a rather unconventional single-atom laser/amplifier and investigated all relevant quantities (such as photon transmission, phase response, threshold, spectrum, photon number and $g^{(2)}$ correlation function). Phonons and the leads under finite bias are established to be the source of relaxation, pumping and pure dephasing mechanisms. We obtain the analytical results for the DC current, photon flux, photon transmission and phase response below and above threshold and these results are benchmarked by exact Master equation simulations. When applicable, we compared our analytical and numerical results to existing experimental data of photon transmission, phase response and non-equilibrium steady state electron current. We believe that these results, both analytical and numerical, are of paramount importance in establishing the optimal parameter choice for future experiments to reach the strong-coupling regime and lasing, as studied before in atomic and solid-state Cavity QED \cite{muller85,kimble,ates}. 


\section{Acknowledgements}

We would like to thank J. R. Petta, Y. Liu, K. Petersson and G. Stehlik for sharing their experimental data and for several enlightening discussions. We thank A. Wacker, M. Vavilov and C. Xu for useful discussions. This work was supported by the US National Science Foundation through the NSF CAREER Grant No. DMR-1151810 and the Eric and Wendy Schmidt Innovative Technology Fund.


%
%
%
%
%
%
%
%
%
%
%

\appendix




\section{Electron Transport}
\label{appA}
In this appendix we analyze transport in the special case when there is at most one electron on the DQD. 


The starting Hamiltonian is: 
\bea
H=H_{DQD}+H_{leads} + H_{leads-DQD}\\
H_{DQD}= \omega_L e_L^\dagger e_L + \omega_R e_R^\dagger e_R + t (e_L^\dagger e_R + e_R^\dagger e_R)\\
H_{leads} = \sum_{kj} \omega_{kj} b_{kj}^\dagger b_{kj} \\
H_{leads-DQD}=\sum_{kj} (t_{kj} b_{kj} e_j ^\dagger +h.c.)
\eea
where $j \in \{L,R\}$,  $e_{L}^\dagger \equiv \ket{L}\bra{0}$ and $e_{R} ^\dagger \equiv \ket{R}\bra{0}$ are the creation operators for an electron on the left and right QDs and $b_{kj}^\dagger = \ket{k, j}\bra{0}$  is the creation operator for an electron in state $k$ in the $j$th reservoir. 

In the Schr\"odinger picture the Born-Markov second order master equation takes the form  \cite{timm08} :
\bea\label{rrr}\label{A5}
\dot{\rho}_{DQD}(t) &=& - i [H_{DQD}, \rho_{DQD}(t)] \\
&-&  \int_0^{\infty} \di \tau \langle [H_{leads-DQD},\nonumber\\&&[H_{leads-DQD}(\tau),\rho_{dqd}(t) \otimes \rho_{leads}]] \rangle_{H_{leads}}\nonumber \eea

where $O(\tau)\equiv e^{i (H_{dqd}+ H_{leads}) \tau } O e^{-i (H_{dqd}+ H_{leads}) \tau } $ is the interaction picture operator $O$ at time $\tau$. 

Therefore, in order to find the master equation we need to know the time dependence of $H_{leads-DQD}$ in the interaction picture. For this we need to know the time dependence of $e_j^\dagger$ and $b_{kj}^\dagger$: 
\bea\label{A6}
e_L^\dagger(t) &=& \cos\left(\frac{\theta}{2}\right)e^{i E_e t } e^\dagger_e- \sin\left(\frac{\theta}{2}\right)e^{i E_g t } e^\dagger_g\nonumber\\ 
e_R^\dagger(t) &=& \sin\left(\frac{\theta}{2}\right)e^{i E_e t } e^\dagger_e + \cos\left(\frac{\theta}{2}\right)e^{i E_g t } e^\dagger_g\nonumber\\ 
b_{kj}^\dagger(t) &=& e^{i \varepsilon_ {kj} t } b_{kj}^\dagger.
\eea
where $e_e^\dagger = \ket{e}\bra{0}$, $e_g^\dagger = \ket{g} \bra{0}$ are the creation operators for electrons on the DQD in the excited and the ground state and $E_{e,g,0}$, the eigenenergies of the DQD are: 
\bea
E_e&=&\frac{\omega_L+\omega_R}{2} +\frac{\Omega}{2},\\
E_g&=&\frac{\omega_L +\omega_R}{2} - \frac{\Omega}{2}, \\
E_0&=&0
\eea
where $\Omega=\sqrt{(\omega_L-\omega_R)^2 + 4 t^2}$. 

To evaluate \ref{rrr} we use the fact that the leads are fermionic reservoirs at zero temperature with the chemical potential $\mu_j$, and ignore the usually trivial effects of Lamb-shift type of terms to obtain: 
\bea\label{A9}
\langle \sum_k b_{kj} \rangle_{H_{leads}}=\langle \sum_k b^\dagger_{kj} \rangle_{H_{leads}} = 0\nonumber\\
\langle \int_0^\infty \di \tau \sum_k t_{kj}^* b^\dagger_{kj}(0) \sum_q t_{qj} b_{qj} (\tau) e^{i \omega_0 \tau}\rangle_{H_{leads}}\nonumber\\
= \int_0^\infty \di \tau \sum_{k,q} t_{kj}^*t_{qj} \langle b^\dagger_{kj}(0) b_{qj} (0)\rangle_{H_{leads}} e^{i (\omega_0-\omega_{qj} ) \tau}\nonumber\\
=  \int_0^\infty \di \tau \sum_{k,q} t_{kj}^*t_{qj} \delta(k,q) \Theta(\mu_j - \omega_{qj}) e^{i (\omega_0-\omega_{qj} ) \tau}\nonumber\\
=  \sum_{k} |t_{kj}|^2  \Theta(\mu_j - \omega_{kj}) \pi \delta(\omega_0-\omega_{kj})\nonumber\\
=\frac{1}{2} J(\omega_0) \Theta (\mu_j-\omega_0)
\eea
where $J_j(\omega) \equiv 2 \pi  \sum_k |t_{kj}|^2 \delta(\omega-\omega_{kj})$ is the spectral density function of the $j$th reservoir and $\Theta(x)$ is the Heaviside step function.

For simplicity we evaluate the special case when $\mu_L>E_e, E_g>\mu_R$ and we assume that the reservoir spectral density function varies slowly between $\omega = E_e$ and $\omega=E_g$ such that we can make the approximation $J_j(E_g) \approx J_j(E_e)$. In this case we introduce \ref{A6} into \ref{A5} and use \ref{A9} to obtain the following master equation:
\bea\label{nonsecularME}
\dot{\rho}_{DQD} &=& - i [H_{DQD} , \rho_{DQD}]\\
& & + \Gamma_L \mathcal{D} \left( \cos\left(\frac{\theta}{2}\right)\ket{e}\bra{0}+ \sin\left(\frac{\theta}{2}\right)\ket{g}\bra{0}\right)\nonumber\\ 
& &+ \Gamma_R \mathcal{D}\left(-\sin\left(\frac{\theta}{2}\right)\ket{0}\bra{e}+\cos\left(\frac{\theta}{2}\right)\ket{0}\bra{g}\right)\nonumber
\eea

We ``secularize" equation \ref{nonsecularME}   as described in  \cite{dumcke1979,keeling13,nazir2012} . In this approximation, the excited and the ground state of the DQD are assumed to interact independently with the lead electrons. Secularization is valid as long as the linewidths of the ground and excited states of the DQD are smaller than the energy difference between them and the two states can be distinguished clearly. When this is not the case (i.e. $\Gamma_{L,R} \ge E_e - E_g$) we expect this approximation to break down. The secularized master equation is:
\bea \label{master}
\dot{\rho}_{DQD} &=& - i [H_{DQD} , \rho_{DQD}]+ \Gamma_L  \cos^2 \left(\frac{\theta}{2}\right)\mathcal{D}(\ket{e}\bra{0})\nonumber\\&+&\Gamma_L  \sin^2 \left(\frac{\theta}{2}\right)\mathcal{D}(\ket{g}\bra{0})\nonumber\\
&+&  \Gamma_R  \sin^2 \left(\frac{\theta}{2}\right)\mathcal{D}(\ket{0}\bra{e})\nonumber\\&+&\Gamma_R  \cos^2 \left(\frac{\theta}{2}\right)\mathcal{D}(\ket{0}\bra{g})\nonumber\\
H_{DQD}&=& \omega_L e_L^\dagger e_L + \omega_R e_R^\dagger e_R + t (e_L^\dagger e_R + e_R^\dagger e_R)
\eea
In the eigenbasis $H_{DQD}$ has the form:
\bea
H_{DQD}&=&\frac{\omega_L+\omega_R}{2} I_2 + \frac{\Omega}{2}\sigma_z\nonumber,\\
&=&-\frac{\omega_L+\omega_R}{2}\ket{0}\bra{0} + \frac{\Omega}{2}\sigma_z
\eea
where where we used the Pauli spin operators in the eigenbasis $\left\{\ket{e},\ket{g}\right\}$ (i.e. $I_2=\ket{e}\bra{e}+ \ket{g} \bra{g}$, $\sigma_z =\ket{e}\bra{e}-\ket{g}\bra{g}$,  $\sigma_+=\ket{e}\bra{g}$ and $\sigma_- = \ket{g}\bra{e}$). 

Since the zero state energy does not enter the dynamics anymore (this energy only mattered in obtaining the rates $\Gamma_{L,R}$) we can ignore the term proportional $\ket{0}\bra{0}$ with impunity to obtain the final Hamiltonian: 
\bea
H_{DQD} = \frac{\Omega}{2}\sigma_z.
\eea
\section{Equations of Motion}
\label{EOMapp}

\begin{figure}[t!]
	\centering
	\includegraphics[]{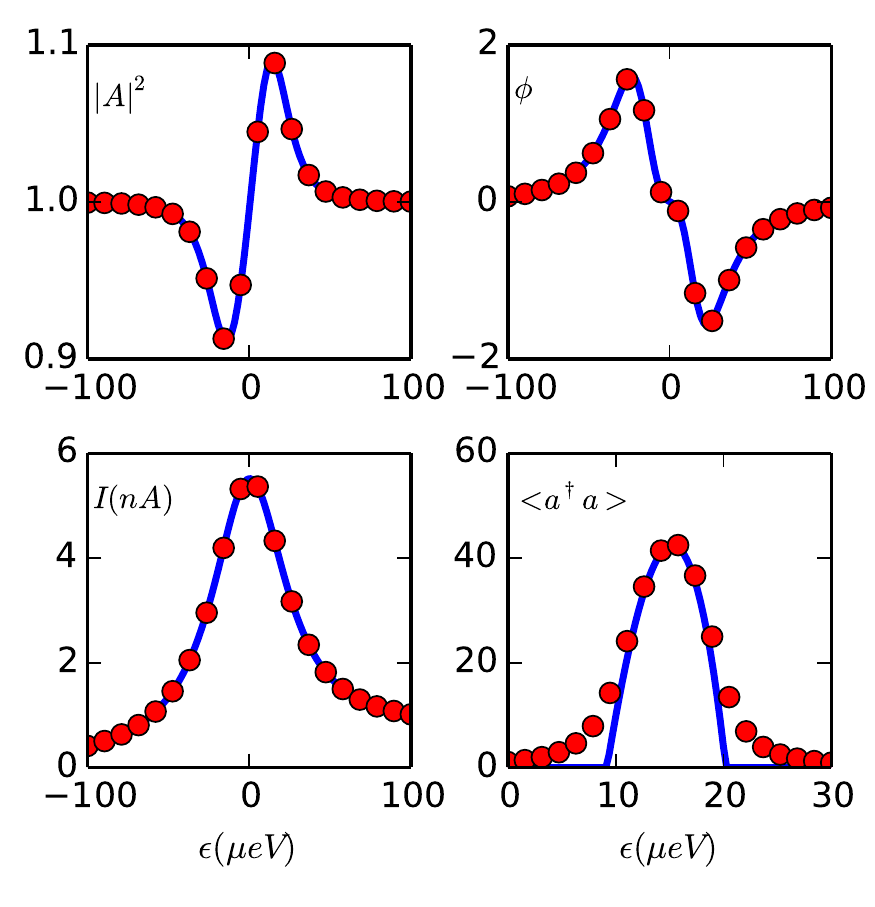}
	\caption{ (top left) Photon transmission versus detuning ($\epsilon$) (top right) Phase response versus detuning (bottom left) Non-equilibrium steady state current versus detuning (bottom right) Cavity photon number versus detuning. Analytics is solid line (blue) while exact numerics is red circles. Parameter values are given in Table \ref{tab:title} except for photon emission (bottom right) where $g_0\rightarrow5.3g_0$ so that we can go above threshold. The discrepancy in photon number exhibited in the bottom right figure is exactly what we would expect from a semiclassical approximation. The analytical results are generated from Eq. \ref{TRANS} (transmission and phase response), Eq. \ref{PHOT} (photon number) and Eq. \ref{CUR} (DC electron current).}
	\label{fig:tran}
\end{figure}

\begin{figure}[t!]
	\centering
	\includegraphics[]{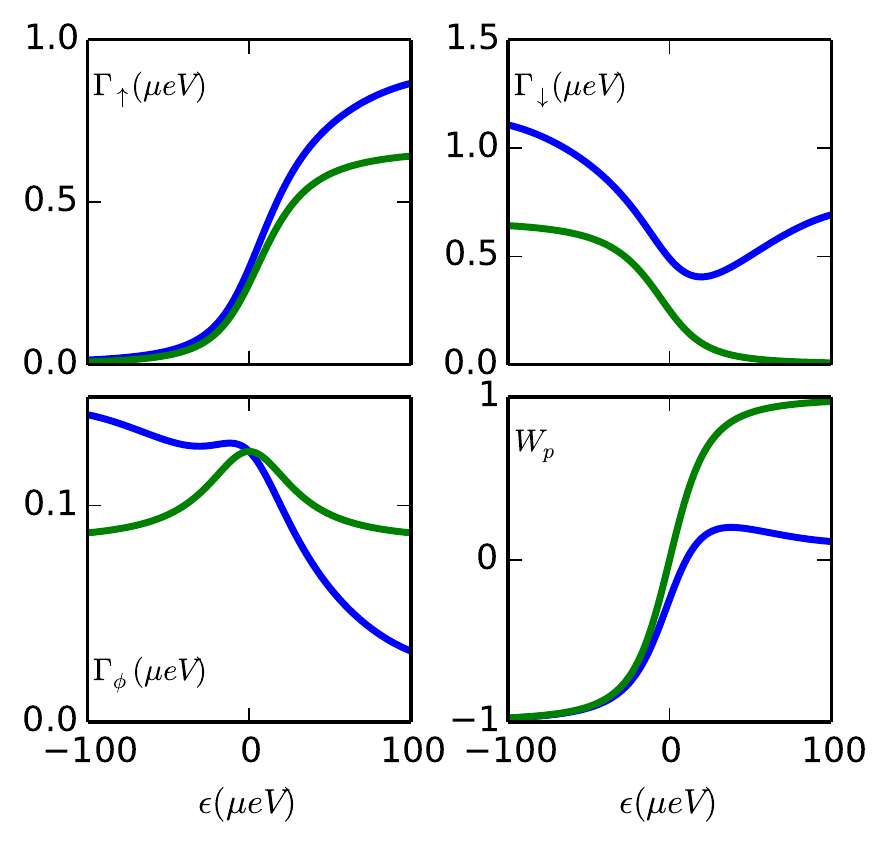}
	\caption{Top left: $\Gamma_\uparrow$, top right $\Gamma_\downarrow$, bottom left $\Gamma_\phi$, bottom right $W_p$. $\Gamma_R=\Gamma_L=1 \mu$eV , the rest of the experimental values are as in Table I. Green is without phonons, blue is with phonons. }
	\label{fig:WP}
\end{figure}
\subsection{Starting Equations of Motion}
We start from the secularized master equation:
\bea
\dot{\rho} &=& - i [H,\rho] + \gamma_\downarrow \mathcal{D}(\sigma_-) +\gamma_\uparrow \mathcal{D}(\sigma_+) +\gamma_\phi\mathcal{D}(\sigma_z) + \kappa D(a)\nonumber \\
& &+ \Gamma_L  \cos^2 \left(\frac{\theta}{2}\right)\mathcal{D}(\ket{e}\bra{0})+\Gamma_L  \sin^2 \left(\frac{\theta}{2}\right)\mathcal{D}(\ket{g}\bra{0})\nonumber\\
& &+  \Gamma_R  \sin^2 \left(\frac{\theta}{2}\right)\mathcal{D}(\ket{0}\bra{e})+\Gamma_R  \cos^2 \left(\frac{\theta}{2}\right)\mathcal{D}(\ket{0}\bra{g})\nonumber,\\
H &=&\frac{\Omega}{2}\sigma_z + g (\sigma_+ a + \sigma_- a^\dagger) \nonumber\\&+& \omega_c a^\dagger a + i\sqrt{\frac{\kappa}{2}} E \cos (\omega_d t) (a^\dagger - a).
\eea

To solve the above problem we use linear response theory. Since $E$ is a small perturbation we ignore the time dependent term $i \sqrt{\kappa/2} E \cos (\omega_d t) (a^\dagger - a)$  at first. We assume that the system will reach a steady state in which the photon electric field and DQD electric dipole oscillate at a frequency $\omega_l$ which we call the laser frequency and will be determined subsequently. Therefore, we look at the equations of motion in a frame rotating  at $\omega_l$: 
\bea\label{B3}
\langle\dot{a}\rangle&=&- i \Delta_{cl} \langle a \rangle - \frac{\kappa}{2}\langle a \rangle - i g \langle\sigma_- \rangle \\
\langle\dot{\sigma}_-\rangle&=& -(i \Delta_{ql} +2 \gamma_\phi + \frac{\gamma_\uparrow+\gamma_\downarrow+\Gamma_R}{2})\langle\sigma_- \rangle\nonumber\\&+& i g \langle a \sigma_z\rangle ,\\
\langle\dot{\sigma}_z\rangle &=& - \left(2 \gamma_\downarrow + \Gamma_R \sin^2 \left(\frac{\theta}{2}\right)\right) \frac{1}{2}(\langle {I}_2\rangle + \langle\sigma_z\rangle) \nonumber\\
& &+\left(2 \gamma_\uparrow+\Gamma_R \cos^2 \left(\frac{\theta}{2} \right) \right)\frac{1}{2}(\langle I_2\rangle - \langle\sigma_z\rangle)  \nonumber\\
& &+2 i g (\langle a^\dagger \sigma_-\rangle - \langle a \sigma_+\rangle)+\Gamma_L \cos \theta \langle\sigma_{0}\rangle \\
\langle\dot{\sigma_{0}}\rangle&=&\Gamma_R \sin^2 \left(\frac{\theta}{2}\right)\frac{1}{2}(\langle {I}_2\rangle +\langle\sigma_z\rangle) \nonumber\\&+& \Gamma_R \cos^2  \left(\frac{\theta}{2}\right)\frac{1}{2}(\langle{I}_2\rangle -\langle \sigma_z\rangle) - \Gamma_L\langle \sigma_{0}\rangle\\
1&=&\langle{I}_2\rangle+\langle\sigma_{0}\rangle.\label{B6}
\eea
where $\sigma_i$ are the usual Pauli spin matrices in the eigenbasis $\{\ket{e}, \ket{g}\}$, $I_2\equiv\ket{e}\bra{e} + \ket{g}\bra{g}$ and $\sigma_0\equiv \ket{0}\bra{0}$. We are only interested in the steady state solutions. 

Integrating out $\sigma_{0}$ we obtain the following equations in the steady state: 
\bea\label{3LSteadyBegin}
0&=&- i \Delta_{cl} \langle a \rangle_{ss} - \frac{\kappa}{2}\langle a \rangle_{ss} - i g \langle\sigma_- \rangle_{ss} ,\\
0&=& -(i \Delta_q + \gamma_d)\langle\sigma_- \rangle_{ss}+ i g \langle a \sigma_z\rangle_{ss} ,\\
0&=& -\gamma_r (\langle I_2\rangle_{ss}+ \langle\sigma_z\rangle_{ss})  \nonumber\\&+&\gamma_p (\langle I_2\rangle_{ss} \nonumber\\&-& \langle\sigma_z\rangle_{ss}) +2 i g (\langle a^\dagger \sigma_-\rangle_{ss} -\langle a \sigma_+\rangle_{ss}) ,\\
\label{3LSteadyEnd}\langle I_2\rangle_{ss} &=& \frac{1+\langle\sigma_z\rangle_{ss} \cos \theta \Gamma_R/(2 \Gamma_L)}{1+\Gamma_R/(2\Gamma_L)} ,
\eea
where $\gamma_r \equiv \gamma_\downarrow+ \Gamma_R \sin^4 \left(\theta/2\right)$, $\gamma_p\equiv \gamma_\uparrow + \Gamma_R \cos^4 \left( \theta/2\right)$ and $\gamma_d \equiv 2\gamma_\phi+(\gamma_\uparrow+\gamma_\downarrow+\Gamma_R)/2$. 

Introducing the last equation into the third equation we obtain: 
\bea
\langle\dot{\sigma}_z \rangle_{ss}&=&0= \gamma(W_p - \langle \sigma_z\rangle_{ss}) \nonumber\\&+&2 i g (\langle a^\dagger \sigma_-\rangle_{ss} -\langle a \sigma_+\rangle_{ss})\\
\gamma&\equiv& \gamma_p+\gamma_r - (\gamma_p-\gamma_r)\frac{\cos\theta \Gamma_R/(2\Gamma_L)}{1+\Gamma_R/(2\Gamma_L)}\\
W_p&\equiv&\frac{1}{\gamma}\frac{\gamma_p-\gamma_r}{1+\Gamma_R/(2\Gamma_L)} .
\eea

So far, no approximation have been made. 

We need to solve the following equations, known as the Maxwell-Bloch equations:
\bea\label{eqe}
0&=& - i \Delta_{cl} \langle a \rangle_{ss} - \frac{\kappa}{2}\langle a \rangle_{ss} - i g \langle\sigma_- \rangle_{ss} ,\\
0&=& -(i \Delta_{ql} + \gamma_d)\langle\sigma_- \rangle_{ss}+ i g \langle a \sigma_z\rangle_{ss} ,\\
0&=& \gamma(W_p - \langle \sigma_z\rangle_{ss}) \nonumber\\&+&2 i g (\langle a^\dagger \sigma_-\rangle_{ss} -\langle a \sigma_+\rangle_{ss}).
\eea
Although the above equations do not contain the third level, the trace of the density matrix is not $1$. This information is contained in the value of $I_2$ which we also eliminated in the above. Therefore, the above equations describe the equations of motion of a two level system, and can be obtained from the following two-level master equation: 
\bea
\label{sal11}
\dot{\rho} &=& - i [H,\rho] \nonumber+ \Gamma_\downarrow \mathcal{D}(\sigma_-)\\& &+ \Gamma_\uparrow \mathcal{D}(\sigma_+) +\Gamma_\phi \mathcal{D}(\sigma_z) + \kappa D(a),\nonumber\\
H &=&\frac{\Omega}{2}\sigma_z + g (\sigma_+ a + \sigma_- a^\dagger) \nonumber\\&+& \omega_c a^\dagger a + i\sqrt{\frac{\kappa}{2}} E \cos (\omega_d t) (a^\dagger - a).
\eea
with effective pumping, relaxation and pure dephasing rates given by 
\bea
\Gamma_\uparrow = \gamma(1+W_p)/2\\
\Gamma_\downarrow = \gamma(1-W_p)/2\\
\Gamma_\phi =( \gamma_d - \gamma/2)/2
\eea

Notice that the two level system reduction is exact when determining the steady state expectation values of operators and correctly describes the time evolution of the photon field when $\kappa \ll \Gamma_{\downarrow,\uparrow,\phi}$, when the DQD will adiabatically follow the photon field. To determine values of experimentally relevant operators, we express the operators in terms of the three level system operators, then use equations \ref{3LSteadyEnd} and \ref{B6} to express the operators in terms of two-level system operators obtained from solving \ref{sal11}.

\subsection{Semiclassical Approximation}\label{sec:semi}
Notice that the system of equations \ref{eqe} is not closed. In fact we can never obtain a closed system of equations using this method. If we were to find the equations of motion for quantities like $\langle a \sigma_z \rangle$ we would obtain equations that depend on other new terms and so on. Since the master equation cannot be solved exactly there are two main approximations that are usually employed at this stage. Firstly, one can do a semi-quantum approximation which assumes $\langle \sigma_z a  \rangle \approx \langle \sigma_z\rangle \langle a \rangle$. An even stronger approximation, which we will use, is the semiclassical approximation which assumes that the cavity field $a$ resembles a classical field. This allows us to factorize operator expectation values such that  $\langle a O \rangle \approx \langle a \rangle \langle O \rangle$ for any DQD operator $O$. In this approximation the equations of motion become:

\bea\label{eqee}
0&=& - i \Delta_{cl} \langle a \rangle_{ss} - \frac{\kappa}{2}\langle a \rangle_{ss} - i g \langle\sigma_- \rangle_{ss} ,\\
0&=& -(i \Delta_{ql} + \gamma_d)\langle\sigma_- \rangle_{ss}+ i g \langle a \rangle_{ss} \langle \sigma_z\rangle_{ss} ,\\
0&=& \gamma(W_p - \langle \sigma_z\rangle_{ss}) \nonumber\\&+&2 i g (\langle a^\dagger \rangle_{ss}\langle  \sigma_-\rangle_{ss} -\langle a \rangle_{ss}\langle \sigma_+\rangle_{ss}).
\eea

The last two equations imply: 
\bea
\langle \sigma_z \rangle_{ss} &=& \frac{W_p}{1 + G |\langle a \rangle_{ss}|^2} ,\\
\langle \sigma_- \rangle_{ss} &=& i g |\langle a \rangle_{ss}| \frac{\langle \sigma_z\rangle_{ss}}{i \Delta_{ql} + \gamma_d} ,
\eea
where $G= 4 g^2 \gamma_d / (\gamma (\gamma_d^2 + \Delta_{ql}^2))$.  
Introducing this back into the EOM for the cavity field we obtain the following equation: 
\bea \label{eq1}
\langle a \rangle_{ss}\left( 1 - \frac{1}{\frac{\kappa}{2} + i \Delta_{cl}}\cdot \frac{g^2}{i \Delta_{ql} + \gamma_d} \cdot\frac{W_p}{1 + G |\langle a \rangle_{ss}|^2} \right) = 0 \nonumber.
\eea

Equating the imaginary parts on both sides of the equation we obtain the so called line pulling formula which determines the lasing frequency $\omega_l$: 
\bea
\omega_l = \omega_c \frac{\gamma_d}{\gamma_d + \kappa/2} + \Omega \frac{\kappa/2}{\kappa/2 + \gamma_d} .
\eea
The above equation implies that the lasing frequency will be "pulled" towards the frequency of the higher quality part of the system. 

Analysing equation \ref{eq1} we see that there is a critical $W_p$ which yields qualitatively different solutions. When $W_p<2 \kappa / (G \gamma)$ only the trivial solution $|\langle a \rangle_{ss}|^2=0$ is allowed. However, when $W_p>2 \kappa / (G \gamma)$ another more interesting solution is allowed. This critical $W_p$ is known as the laser threshold $W_{th} \equiv 2 \kappa / (G \gamma)$. 


Therefore we express the cavity field intensity as:
\bea\label{aaa}
|\langle a \rangle_{ss}|^2 &=&
\begin{cases} 0, & W_p <W_{th},\\  
\frac{1}{G}\left(\frac{W_p}{W_{th}}- 1\right), & W_p >W_{th}.
\end{cases}.
\eea

The steady state inversion is: 
\bea\label{iii}
\langle \sigma_z \rangle_{ss}=
\begin{cases} W_p, & W_p <W_{th}\\ 
 	W_{th}, & W_p >W_{th} 
\end{cases} .
\eea

We can also obtain the steady state current:
\bea
I_{ss} \equiv \Gamma_R \rho_{RR} &=& \frac{\Gamma_R}{2(1 + \Gamma_R/(2\Gamma_L))} (1 - \langle \sigma_z \rangle_{ss} \cos\theta) \nonumber\\&+& \Gamma_R \sin\theta \Re (\langle \sigma_- \rangle_{ss})\\
 &=&  \frac{\Gamma_R}{2(1 + \Gamma_R/(2\Gamma_L))} (1 - \langle \sigma_z \rangle_{ss} \cos\theta) \nonumber,
\eea
where we used the fact that $\langle \sigma_- \rangle_{ss} = 0$ due to phase diffusion as we subsequently show in \ref{alzero}.
\subsection{Transmission}
\label{appT}
We have found the expressions for the cavity field intensity and inversion. At this point we reintroduce the classical perturbative probing field $E \cos(\omega_d t)$ into the initial equations and calculate the other system operators keeping in mind that inversion and field intensity are fixed due to nonperturbative processes. According to linear response theory the perturbation $E$ will introduce a oscillatory perturbation rotating at the driving frequency $\omega_d$ such that the steady state cavity field is:
\bea \label{Ans}
 \lim_{t \to \infty}a(t) = a_l e^{-i \omega_l t} +a_d e^{-i \omega_d t }
 \eea
where $a_l$ is the field operator in the absence of the driving field $E$ while $a_d$ is a small perturbation induced by the driving field. The DQD coherence $\sigma_-$ will be similarly affected such that: 
\bea
\lim_{t \to \infty}\sigma_-(t) = \sigma_-^l e^{-i \omega_l t} +\sigma_-^d e^{-i \omega_d t }
\eea

In this case the equations of motion for $\langle a_l \rangle $ are: 
\bea
\langle \dot {a}_l \rangle_{ss} &=& 0 = - i \Delta_{cl} \langle a_l \rangle_{ss} - \frac{\kappa}{2}\langle a_l  \rangle_{ss}\nonumber\\ & &- i g \langle\sigma_-^l \rangle_{ss}\\
\langle \dot{\sigma}_-^l\rangle_{ss}  &=&0= -(i \Delta_{ql} + \gamma_d)\langle\sigma_-^l \rangle_{ss}\nonumber\\ & &+ i g \langle a_l  \rangle_{ss} \langle \sigma_z\rangle_{ss} 
\eea
where the inversion $\langle \sigma_z\rangle_{ss}$ was calculated in \ref{iii}. It's easy to solve the above equations of motion and obtain: 
\bea\label{alzero}
\langle  a_l  \rangle_{ss} = \langle \sigma_-^l \rangle_{ss} =0
\eea
This effect is known as phase diffusion\cite{cmb}. 
The equations of motion for $\langle a_d \rangle $ are: 
\bea
\langle \dot {a}_d \rangle_{ss} &=& 0 = - i \Delta_{cd} \langle a_d \rangle_{ss} - \frac{\kappa}{2}\langle  a_d  \rangle_{ss}\nonumber \\& &- i g \langle\sigma_-^d \rangle_{ss}+\sqrt{ \kappa/8} E\\
\nonumber\langle \dot{\sigma}_-^d\rangle_{ss}&=&0= -(i \Delta_{qd} + \gamma_d)\langle\sigma_- ^d\rangle_{ss}+ \\& &i g \langle  a_d   \rangle_{ss} \langle \sigma_z\rangle_{ss} 
\eea
where, again, the inversion $\langle \sigma_z\rangle_{ss}$ was calculated in \ref{iii}  and is not influenced by the perturbation $E$. By defining the observable $A\equiv \sqrt{2 \kappa} \langle a_d \rangle_{ss} / E$ we can solve the above to obtain: 
\bea\label{A}
A = \frac{\kappa/2 }{i \Delta_{cd}+\kappa/2 - g^2 \langle \sigma_z\rangle_{ss}/(i \Delta_{qd} + \gamma_d)}
\eea
Notice that $|A|^2$ gives the transmission coefficient of a microwave of strength $E$ through transmission line resonator, while $\phi = \mathrm{arctan}[\I(A)/\R(A)]$ gives the phase shift of the microwave. Although the above formula looks simple it has a complicated dependence on detuning since $g$, $\Delta_{qd}$, $\Delta_{cd}$,$\omega_d$, and $\sigma_z$ are all functions of detuning $\epsilon$.



\bibliographystyle{apsrev}
\bibliography{references}

\end{document}